\documentclass[11pt]{amsart}

\usepackage[T1]{fontenc}
\usepackage{lmodern}
\usepackage[utf8]{inputenc}
\usepackage[english]{babel}
\usepackage{microtype}
\usepackage{mathtools,amssymb,amsthm}
\usepackage{booktabs}
\usepackage{tabularx}
\usepackage{enumitem}
\usepackage[left=1in,right=1in,top=1in,bottom=1in]{geometry}
\usepackage{booktabs}
\usepackage{array}
\usepackage[colorlinks=true,linkcolor=blue,citecolor=blue,urlcolor=blue]{hyperref}

\usepackage{xcolor}
\usepackage{listings}

\lstdefinestyle{Rstyle}{
  language=R,
  basicstyle=\ttfamily\footnotesize,
  keywordstyle=\color{blue!60!black},
  commentstyle=\color{gray}\itshape,
  stringstyle=\color{green!40!black},
  frame=single, rulecolor=\color{black!30},
  breaklines=true, showstringspaces=false,
  columns=fullflexible, keepspaces=true
}
\newtheorem{lemma}{Lemma}[section]
\theoremstyle{definition}

\newtheorem{example}[lemma]{Example}
\theoremstyle{remark}

\newcommand{\PP}{\mathbb{P}}

\newcommand{\1}{\mathbf{1}}
\newcommand{\RU}{\operatorname{RU}}
\newcommand{\TP}{\operatorname{TP}}
\newcommand{\FN}{\operatorname{FN}}
\newcommand{\FP}{\operatorname{FP}}
\newcommand{\NB}{\operatorname{NB}}

\begin{document}

\title{Interpreting relative utility for probabilistic predictions}

\author[Linard Hoessly]{Linard Hoessly}
\address{Data Center of the Swiss Transplant Cohort Study, University of Basel \& University Hospital Basel, Basel, Switzerland}
\email{linard.hoessly@hotmail.com}

\date{}
\keywords{Relative utility, net benefit, decision curve analysis, probabilistic prediction, perfect prediction, perfect information, clinical utility}

\begin{abstract}
At a fixed threshold, relative utility (RU) measures the net-benefit gain of a
prediction model over the better of treat-all and treat-none relative to the
corresponding gain under perfect outcome classification. We illustrate that RU
equal to 1 therefore represents perfect outcome classification at that fixed threshold, not perfect
probabilistic prediction. However, even when every predicted probability equals the
true probability, observed RU can equal 0. In a simple constant-risk setting,
this occurs with probability approaching 1 as the sample size increases.
Consequently, the distance from observed RU to 1 should not in general be
interpreted as improvement achievable by a better prediction for binary probabilities.
\end{abstract}

\maketitle

\section{Introduction}\label{sec:intro}

Clinical prediction models are commonly evaluated in terms of discrimination,
calibration, probabilistic accuracy, and clinical utility
\cite{Steyerberg2010_scaled_2,Hoessly_cje_key2026}. Decision curve analysis and
net benefit(NB) evaluate the consequences of threshold-based decisions
\cite{Vickers2006,Vickers2016}.

Relative utility (RU) was introduced in ~\cite{Baker2009JRSSA} as a decision-analytic measure of prediction
performance. Here we consider its threshold-specific NB
representation \cite{VanCalster2013}. At a threshold \(t\in(0,1)\), a model that
treats individuals with \(p_i\ge t\) has relative utility
\begin{equation}\label{eq:RUdef}
\RU(t)=
\frac{\NB(t)-\max\{\NB_{\mathrm{none}}(t),\,\NB_{\mathrm{all}}(t)\}}
{\NB_{\mathrm{oracle}}(t)-\max\{\NB_{\mathrm{none}}(t),\,\NB_{\mathrm{all}}(t)\}}.
\end{equation}
where $NB(t)
=
\frac{TP(t)}{n}
-
\frac{t}{1-t}\frac{FP(t)}{n}$ with \(\operatorname{TP}(t)\) and \(\operatorname{FP}(t)\) the numbers of true positives and false positives obtained by classifying individuals with predicted probability at least \(t\) as positive, 
\(\NB_{\mathrm{none}}(t)=0\), \(\NB_{\mathrm{all}}(t)=(\bar y-t)/(1-t)\)
and \(\NB_{\mathrm{oracle}}(t)=\bar y\), with \(\bar y\) the observed event rate. Writing \(w=t/(1-t)\), the denominator of \eqref{eq:RUdef} equals
$
\min\{\bar y,\,w(1-\bar y)\}.
$
The denominator is strictly positive if and only if \(0<\bar y<1\). When
\(\bar y\in\{0,1\}\), the denominator is zero and RU is undefined.

The oracle makes no false-positive and no false-negative decisions, so its net
benefit is at least that of any decision rule.  Thus RU is 0 when the model attains the net benefit of the better default
strategy and 1 when it attains the oracle net benefit, but it can also be
negative. Attaining the oracle net benefit requires no false-positive or
false-negative decisions, so RU equals 1 exactly when the observed outcomes
are classified perfectly at the corresponding threshold.

In real data we have predicted probabilities from a clinical prediction model and
observe realised outcomes, but not the individual true probabilities \cite{steyerberg2019clinical}. Writing \(p_i\) for the prediction, \(q_i\) for the unobservable true probability and \(y_i\) for the realised outcome, perfect probabilistic prediction means \(p_i=q_i\), not \(p_i=y_i\); the outcome remains random whenever \(0<q_i<1\). This distinction is central to interpreting the Brier score \cite{Hoessly2026}, and we take the same view here.

\cite{Baker2009JRSSA} described relative utility curves as
gauging the potential for better prediction and for improved performance with
better prediction models. \cite{VanCalster2013} similarly
described RU as the proportion of the possible improvement over the default
strategy captured by the model.

The upper benchmark itself was explicitly linked to perfect clinical
information \cite{Baker2009JRSSA} and has subsequently often been called
``perfect prediction''
\cite{Baker2009JNCI,Baker2014,Baker2019,Binuya2022}.Taken together, these descriptions can naturally suggest that the distance
from observed RU to 1 reflects performance that might be gained by improving
the prediction model. For probabilistic prediction, however, this
interpretation requires qualification. These descriptions are compatible when prediction is understood as outcome
classification. For probabilistic prediction, however, a perfect model reports
the true conditional risks, whereas perfect clinical information reveals the
realised outcomes. This affects the interpretation of both the RU upper
benchmark and the distance to it.

The consequences are not subtle, as can be seen from the following example.
\begin{example}\label{extr_ex}
Suppose five patients have each a true event risk of \(20\%\), and corresponding model
predictions of \(20\%\) for each patient. So the model predicts every patient's
true risk exactly. Suppose that one of the five patients experiences the
event, say the first patient.

At a treatment threshold of \(10\%\), the model treats all five patients and
therefore makes exactly the same decisions as treat-all. Since the observed
event rate is \(20\%\), $\bar y=0.2$ and RU is defined, treat-all is the better default strategy and $NB(t)=\NB_{\mathrm{all}}(t)$, hence
\[
\RU(t)=0.
\]

The oracle, in contrast, knows which patient experiences the event and treats
only that patient, giving $\NB_{\mathrm{oracle}}(t)=0.2$. Thus RU is \(0\) even though the probability predictions
are perfect. The gap to 1 remains despite the absence of any probability-prediction error.
Figure~\ref{fig:perfect_prediction_ru} illustrates this distinction. Appendix~\ref{app:constant} strengthens this observation by showing that, in a simple
constant-risk setting, this occurs with probability approaching one as the sample
size increases. 
\end{example}

\begin{figure}[ht]
\centering

\[
\text{True risk = predicted risk = }20\%\text{ for every patient},
\qquad
t=10\%.
\]

\vspace{0.5em}

\renewcommand{\arraystretch}{1.5}
\begin{tabular}{
    >{\raggedright\arraybackslash}p{0.27\textwidth}
    *{5}{>{\centering\arraybackslash}p{0.105\textwidth}}
}
\toprule
 & Patient 1 & Patient 2 & Patient 3 & Patient 4 & Patient 5 \\
\midrule

Realised outcome
    & Event
    & No event
    & No event
    & No event
    & No event \\

Perfect probability model
(\(\RU=0\))
    & Treat
    & Treat
    & Treat
    & Treat
    & Treat \\

Outcome oracle
(\(\RU=1\))
    & Treat
    & No treatment
    & No treatment
    & No treatment
    & No treatment \\

\bottomrule
\end{tabular}

\caption{
A perfect probability model knows that each patient has a \(20\%\) risk, but
not which patient will experience the event. At a \(10\%\) threshold it
therefore treats everyone, just like treat-all, and has RU \(=0\). The outcome
oracle knows the realised outcomes and has RU \(=1\).
}
\label{fig:perfect_prediction_ru}
\end{figure}
This value of zero is meaningful from a decision perspective: at this
threshold, perfect knowledge of the patients' risks does not change any
decision relative to treat-all. The remaining distance to RU \(=1\) is
therefore not due to error in the predicted probabilities. Similarly, any prediction with $\tilde{p} \in[0.1,1]\times [0,0.1)^4$, i.e. where the threshold-based classification corresponds to the final outcome gives $RU(t)=1$ at $10\%$. 

The example illustrates the main point of this paper: the RU upper benchmark
is defined by the net benefit of an oracle that knows the realised outcomes. Even when every predicted probability equals the true probability, observed
relative utility can equal zero. The
probability of this occurring can approach one as the sample size increases in a simple constant-risk setting (Appendix~\ref{app:constant}). Consequently, the distance to 1 should not in general be interpreted as
potential improvement achievable by a better probability model.

We discuss three consequences for the
interpretation of RU.
\subsection*{Acknowledgements}
We thank Matthew Parry, Tinh-Hai Collet, Lucia de Andres, and Julien Vionnet for helpful discussions and feedback.
\subsection*{AI use}
During the preparation of this manuscript, we used  GPT-4/5
 for minor language edits as well as for latex support aiming to enhance readability, and elicit.com for literature search. After
using it, we reviewed and edited the content as needed
and take full responsibility for its content.
\section{Notation and terminology}
\label{sec:notation}

For \(n\) individuals, let \(p_i\in[0,1]\) denote the predicted probability,
\(y_i\in\{0,1\}\) the realised outcome, and
$q_i$
the true probability. The \(q_i\) are generally unobservable.

Perfect probabilistic prediction means that
$
p_i=q_i
$ for all $i$,
whereas perfect outcome information means knowing the realised outcomes
\(y_i\) in advance. For an individual with \(0<q_i<1\), the outcome remains uncertain even when
its probability is known exactly.

When needed, we write
\[
\RU_y(p;t)
\]
for the observed relative utility calculated from the realised outcomes.
\section{Three implications for interpreting relative utility}
We focus on the interpretation of RU. Mathematical derivations
supporting the results below are provided in Appendix~\ref{app:derivations}.
\subsection{The upper benchmark is not perfect probabilistic prediction}

The RU upper benchmark has commonly been described as ``perfect prediction''
\cite{Baker2009JNCI,Baker2014,Baker2019,Binuya2022}. In the original
decision-analytic construction, the reference strategy is based on perfect
clinical information and therefore makes no false-positive or false-negative
decisions. Observed RU equals 1 exactly when the model likewise produces no
false positives and no false negatives at the corresponding threshold.

Perfect probabilistic prediction, \(p_i=q_i\), does not in general guarantee
this.
\subsection{The distance to 1 is not generally potential model improvement}

Relative utility has been described as gauging the potential for better
prediction \cite{Baker2009JRSSA} and as the proportion of the possible
improvement over the default strategy captured by the model
\cite{VanCalster2013}. For probabilistic prediction, this interpretation
requires qualification. Example~\ref{extr_ex} shows the issue in its simplest
form: RU can equal 0 even when every predicted probability equals the true
probability.

A useful way to make this distinction explicit is to compare observed RU with
the RU that the true probabilities would have produced in the same realised
sample. Writing \(\RU_y(p;t)\) for the observed RU of the actual predictions
and \(\RU_y(q;t)\) for that obtained from the true probabilities, simply adding
and subtracting \(\RU_y(q;t)\) gives
\begin{equation}
1-\RU_y(p;t)
=
\{1-\RU_y(q;t)\}
+
\{\RU_y(q;t)-\RU_y(p;t)\}.
\end{equation}
The first term is the distance to the outcome oracle that remains under
perfect probabilistic prediction in that sample. The second vanishes when
\(p=q\), but need not be positive in a particular realised sample. Thus even
eliminating probability-estimation error need not eliminate the distance to
RU \(=1\).

This does not imply that the gap can never be reduced by better prediction.
In practice, additional predictors may provide genuinely new information about the outcome
and thereby move predictions towards true probabilities. However, the
distance to RU \(=1\) cannot generally be interpreted as improvement
recoverable by more accurately estimating the sameprobabilities.
\subsection{RU expresses net benefit on an outcome-oracle scale}

At a fixed threshold, RU is a normalization of observed net benefit relative
to two reference strategies. Writing
\[
\NB_{\mathrm{default}}(t)
=
\max\{\NB_{\mathrm{none}}(t),\NB_{\mathrm{all}}(t)\},
\]
we have
\[
\RU_y(p;t)
=
\frac{
\NB_y(p;t)-\NB_{\mathrm{default}}(t)
}{
\NB_{\mathrm{oracle}}(t)-\NB_{\mathrm{default}}(t)
}.
\]

Thus the better default is assigned RU \(=0\), whereas the outcome oracle is
assigned RU \(=1\). For \(0\leq\RU\leq1\), RU locates the model's observed net
benefit between these two reference points. For example, RU \(=0.3\) means
that the model's net benefit lies 30\% of the way from the better default to
the outcome oracle on this scale.

The denominator is the gain achieved by the outcome oracle, not the gain
obtainable by predicting the conditional risks perfectly. Hence RU \(=0.3\)
does not mean that the model has achieved 30\% of the performance attainable
by a perfect probability model, nor that the remaining 70\% is recoverable by
improving the probability predictions.

\section{Discussion}

The issue identified here is not an error in the definition of relative
utility. The original work explicitly linked its upper benchmark to perfect
clinical information \cite{Baker2009JRSSA}. The ambiguity arises because this
benchmark has also been described as ``perfect prediction'', and RU as
indicating possible improvement relative to it
\cite{Baker2009JRSSA,Baker2009JNCI,VanCalster2013}. For probabilistic
prediction, knowing the true conditional risks and knowing the realised
outcomes are different.

This distinction clarifies the interpretation of RU. At a fixed threshold, RU
expresses observed net benefit relative to the better default strategy and an
outcome oracle. Perfect probabilistic prediction need not attain that upper
benchmark, and the distance to RU \(=1\) is therefore not generally a measure
of probability-estimation error or of the improvement obtainable by more
accurately estimating the same conditional risks.
\section{Conclusions}

Relative utility equal to 1 represents perfect outcome classification at that threshold, not
perfect probabilistic prediction. Even when every predicted probability equals
the true probability, observed RU can equal zero, and in a simple constant-risk
setting this occurs with probability approaching one. RU should therefore be
interpreted as a normalization of decision performance relative to the better
default strategy and an outcome oracle, not as a measure of probabilistic
accuracy or remaining potential for improving the same conditional risks.

\bibliographystyle{plain}
\bibliography{pred_references}

\appendix
\section{Technical derivations}
\label{app:derivations}

Throughout, fix a threshold \(t\in(0,1)\) and write
\[
w=\frac{t}{1-t},
\qquad
d_i=\1\{p_i\ge t\},
\qquad
d_i^q=\1\{q_i\ge t\}.
\]

\subsection{Observed relative utility}
\label{app:observed}

Let
\[
\bar y=\frac{1}{n}\sum_{i=1}^n y_i.
\]
The model net benefit is
\[
\NB_y(p;t)=\frac{\TP(t)-w\FP(t)}{n},
\]
whereas the oracle net benefit is \(\bar y\). The better default strategy has
net benefit
\[
\max\left\{0,\frac{\bar y-t}{1-t}\right\}.
\]
Correspondingly, when \(0<\bar y<1\), the denominator of relative utility is
\[
D_y
=
\bar y-
\max\left\{0,\frac{\bar y-t}{1-t}\right\}
=
\min\{\bar y,w(1-\bar y)\}.
\]

Since \(\TP(t)+\FN(t)=n\bar y\),
\begin{align}
1-\RU_y(p;t)
&=
\frac{\bar y-\NB_y(p;t)}{D_y} \notag\\
&=
\frac{\FN(t)+w\FP(t)}{nD_y}.
\label{eq:obs_loss}
\end{align}

Hence it follows that
\[
\RU_y(p;t)=1
\quad\Longleftrightarrow\quad
\FN(t)=\FP(t)=0.
\]
Thus observed RU equals 1 exactly when the realised outcomes are classified
perfectly at the corresponding threshold.

Equation~\eqref{eq:obs_loss} also shows that, at a fixed threshold,
\(\RU_y(p;t)\) depends on the predicted probabilities only through the
classification \(d_i=\1\{p_i\ge t\}\).

\subsection{Observed RU under perfect probabilistic prediction}\label{app:perfect_prob}

Let \(\RU_y(q;t)\) denote the observed RU obtained when the true probabilities
\(q_i\) are used as predictions in the same realised sample. Adding and
subtracting this quantity gives
\begin{equation}
1-\RU_y(p;t)
=
\{1-\RU_y(q;t)\}
+
\{\RU_y(q;t)-\RU_y(p;t)\}.
\label{eq:obsdecomp}
\end{equation}

The first term is the distance to the oracle that would remain if the true
probabilities were known exactly. The second compares the actual model with
the decisions induced by the true probabilities in that particular realised
sample.

More explicitly,
\begin{align}
\RU_y(q;t)-\RU_y(p;t)
&=
\frac{\NB_y(q;t)-\NB_y(p;t)}{D_y} \notag\\
&=
\frac{
\sum_{i=1}^n
(d_i^q-d_i)(y_i-t)
}{
n(1-t)D_y
}.
\label{eq:obs_model_difference}
\end{align}

Unlike an expected decision-theoretic regret, the quantity in
\eqref{eq:obs_model_difference} need not be non-negative. The true
probabilities determine the optimal threshold-based decisions in expectation,
but an alternative classification may by chance agree more closely with the
realised outcomes.

\subsection{Derivation for Example~\ref{extr_ex}}
\label{app:example}

In Example~\ref{extr_ex}, \(n=5\), \(p_i=q_i=0.2\), \(t=0.1\), and
\(\bar y=0.2\). The perfect probability model treats everyone, so
\[
\NB_y(p;t)=\NB_{\mathrm{all}}(t)
=\frac{0.2-0.1}{0.9}.
\]
Treat-all is the better default strategy and therefore
\[
\RU_y(p;t)=0.
\]

\subsection{Perfect probabilities with observed RU zero}
\label{app:constant}

Suppose \(p_i=q_i=q\) for all \(i\), where \(t<q<1\), and let
\(Y_1,\ldots,Y_n\) be independent Bernoulli variables with probability \(q\).
Since \(q>t\), the perfect probabilistic model treats everyone and its NB is equal to \(NB_{\mathrm{all}}(t) \).

Whenever
\[
t\le\bar Y<1,
\]
we have \(NB_{\mathrm{all}}(t)\geq NB_{\mathrm{none}}(t)\), and the
model has the same net benefit as treat-all. Hence
\[
\RU(t)=0.
\]
By the law of large numbers \cite{georgii2008stochastics}, \(\bar Y\to q>t\) in probability, which excludes the event $\{\bar Y=0\}$, while
\[
\PP(\bar Y=1)=q^n\xrightarrow{n\to\infty}0.
\]
Therefore
\[
\PP\{\RU(t)\text{ is defined and equals }0\}\xrightarrow{n\to\infty}1.
\]

Thus observed RU equals zero with probability approaching one even though
every predicted probability equals the true probability.

\end{document}